\documentstyle[aps,twocolumn,prl,epsf]{revtex}
\begin{document}
\draft\twocolumn[\hsize\textwidth\columnwidth\hsize
\csname@twocolumnfalse\endcsname
\title{Generalized Zipf's Law in proportional voting processes}

\author{M. L. Lyra$^1$, U. M. S Costa$^{1,2}$, R. N. Costa Filho$^2$,
and J. S. Andrade Jr.$^2$}

\address{$^1$Departamento de F\'{\i}sica,
Universidade Federal de Alagoas, 57072-970 Macei\'o, Alagoas, Brazil\\
$^2$Departamento de F\'{\i}sica, Universidade Federal do
Cear\'a, 60455-760, Fortaleza, Cear\'a, Brazil}

\maketitle

\begin{abstract}
Voting data from city-councillors, state and federal deputies
elections are analyzed and considered as a response function of a
social system with underlying dynamics leading to complex
behavior. The voting results from the last two general Brazilian
elections held in 1998 and 2000 are then used as representative data
sets. We show that the voting distributions follow a generalized Zipf's
Law which has been recently proposed within a nonextensive statistics
framework. Moreover, the voting distribution for city-councillors is
clearly distinct from those of state and federal deputies in the sense
that the latter depicts a higher degree of nonextensivity. We relate
this finding with the different degrees of complexity corresponding to
local and non-local voting processes.
\end{abstract}
\vskip2pc]

\section{Introduction}

The quantitative analysis of data generated by complex systems is a
standard problem in statistical physics that encounters applications
is several branches of natural and social sciences. Quite frequently,
the statistical analysis reveal the emergence of power-law
distributions as a signature, indicating that these systems
self-organize into a critical state with no characteristic length or
time scales. Recently, several scale-free phenomena have been reported
to occur in social sciences, as for example, in city morfologies 
\cite{citymorfologies}, economic activity 
\cite{econophysics1,econophysics2} and linguistics 
\cite{zipf,linguistics}, 
among others.

One of the most fundamental processes in democratic societies concerns
to elections. The future development of a human democratic
organization is strongly dependent on the results of a series of such
processes. Therefore, several efforts have been made to understand the
way people make their choices \cite{vote} and what social, economic and
political features are relevant in such a way to influence, or even
determine, the general outcome of a particular electoral process.
From the statistical physics point of view, the result of an electoral
process can be considered as a response function of an interacting
and open many-particle system governed by an intricate (and even unknown)
internal dynamics. After the seminal work of Bak {\em et al.}
\cite{baksoc}, there has been a general believe that these systems
usually evolve towards a critical state whose response functions
depict no characteristic scales. As such, the distribution function
related to any stochastic quantity of interest must display asymptotic
power-laws. Further, these power-law distributions must be quite
robust depending only on a few general requirements. In equilibrium
critical phenomena the relevant features are related to the symmetry of
the local variable, the space dimensionality and the range of
interactions. However, in self-organized dynamical systems further
symmetries may become relevant. Therefore, a quantitative study of the
distribution functions obtained from distinct electoral processes and
their consequent classification is a fundamental step towards a better
understanding of the underlying dynamics.

In 1999, Costa Filho {\em et al.} reported an extensive analysis of
the proportional elections held in Brazil in October 1998
\cite{costafilho}. They have shown that the distribution of votes
among candidates for state deputies follows a Zipf-like \cite{zipf}
power-law, $N(v)\propto v^{-\alpha}$, with $\alpha \simeq 1.0$,
extending over two orders of magnitude. Here $N$ is the number of
candidates that received the fraction of votes $v$. This fact reveals
that the electoral process indeed displays features of a
scale-invariant phenomena. Further they observed that such scaling law
is quite robust, being practically the same for different states with
large social and economic discrepancies. This therefore suggests that
the reported distribution is characteristic of a particular
universality class associated with the state deputies election
process. 

The appearance of the $1/v$ distribution has been interpreted in
Ref.~\cite{costafilho} with the assumption that the success of the
candidates determining $N(v)$ may be described by a typical
multiplicative process \cite{west}. In this way, the voting fraction
of a candidate, $v$, can be viewed as a ``grand process'' depending on
the successful completion of a number $n$ of independent
``subprocesses". These factors should be intrinsically related to the
attributes and/or abilities of the candidate to persuade voters and
obtain votes more effectively. As a result, one could then associate
with each candidate the probability $p_i$ of performing the subprocess
$i$ among voters, so that his or her voting fraction would be $v=p_1
p_2 \cdots p_n$. Considering that $p_i$ are independent positive
random variables and $n$ is sufficiently large, we readily obtain from
the central limit theorem that the distribution of $v$ should be
approximately log-normal. In addition, if the dispersion of the
log-normal is sufficiently large, one can observe a $1/v$ type of
distribution, over a wide range of random variable values \cite{west}.

In this work we extend the quantitative study of electoral
distribution functions by including an analysis of the results of the
proportional elections in Brazil held in 2000. That year, all states
in the country voted to choose local city-councillors. For comparison,
we also include an analysis of the congressmen and state deputies
elections of 1998. We find that the scaling behavior of vote
distributions for state deputies and congressmen are the same within
some statistical accuracy but quite distinct from that obtained for
the local voting process for city-councillors. Further, we show that
deviations from the power-law behavior can be well described by a
generalized Zipf's law which has been recently introduced in
connection with the nonextensive statistical mechanics of Tsallis
\cite{tsallis}.

\section{Generalized Zipf's law}

Data from the Brazilian general elections held in 2000 were made
available through the web site of the Brazilian Federal Electoral
Court \cite{tse}. All Brazilian cities held elections for mayor and
city-councillor. We consider the results from the 15 largest (most
populated) cities for the positions of city-councillors. The average
number of votes per candidate is approximately 500. This number is
significantly smaller than the corresponding ones for the elections 
of state and federal deputies.

For each city, we normalize the votes of each candidate by the total
number of voters and construct histograms to give the number of
candidates, $N$, which received a certain fraction of votes, $v$. All
15 histograms present a common form within the available statistical
accuracy. Data from the 15 histograms were grouped in a unique
histogram which is shown as a log-log plot in Fig.~1. The
resulting curve follows very closely the extended form of the Zipf's
law \cite{mandelbrot}
\begin{equation}
N(v)= \frac{A}{(1+Cv)^{\alpha}}~,
\end{equation}
where $A$ is a normalization constant, $C$ governs the crossover
between the initial plateau and the power-law regime characterized 
by the exponent $\alpha$.

The above law has been deduced within the nonextensive thermodynamic
formalism proposed by Tsallis \cite{tsallis} by means of heuristic
arguments based on the fractal structure of symbolic sequences with
long-range correlations \cite{denisov}. Indeed, the properties of
non-linear maps at the chaos threshold have been widely used to better
understand some features related to interacting many-particle systems
presenting complex behavior. In particular, the fast exponential
dynamics observed in chaotic regimes can be reproduced by noticing
that the phase-space volume visited by the map follows a simple linear
differential equation in the form
\begin{equation}
\frac{dW(t)}{dt}=\lambda W~,
\end{equation}
where lambda is the Lyapunov coefficient. This behavior is replaced by
a slower power-law dynamics at critical points where long-range
correlations develop. In this case, $W(t)$ satisfies the non-linear
differential equation
\begin{equation}
\frac{dW(t)}{dt}=\lambda_qW^q~,
\end{equation}
where the exponent $q$ is related to the degree of nonextensivity
induced by the underlying long-range correlations. This is the same
parameter characterizing the proper nonextensive Tsallis entropy
$S_q$ which evolves at a constant rate \cite{lyra}.
A similar generalization can be done to characterize the possible
behaviors of the distribution $N(v)$ \cite{linguistics}. Its
exponential decay, expected to hold for uncorrelated processes, can be
directly obtained by assuming that $N(v)$ satisfies the first-order
linear differential equation
\begin{equation} \label{dif}
\frac{dN(v)}{dv}=-\lambda N~.
\end{equation}
Further, for long-range correlated processes, the extended Zipf's law
follows from the non-linear generalization
\begin{equation}
\frac{dN(v)}{dv}=-\lambda_q N^q
\end{equation}
which provides
\begin{equation} \label{Zipf}
N(v)=\frac{N(0)}{[1+(q-1)\lambda v]^{1/(q-1)}}~,
\end{equation}
where the limit of $q\rightarrow 1$ recovers the usual exponential
form.

The best fit of $N(v)$ for the city-councillors election to
Eq.~(\ref{Zipf}) is shown as a solid line in Fig.~1 and provides
$\alpha=1/(q-1)=2.63$ ($q=1.38$). The exponent is
substantially larger than the one reported to hold for the state
deputies elections of 1998 \cite{costafilho}. In Fig.~2 we show $N(v)$
as computed from data of the state representatives election in the 15
states corresponding to the capitals considered above. This figure is
similar to the one reported in Ref.~\cite{costafilho}. After the
initial plateau, a power-law scaling regime sets up with an exponent
close to $-1.0$. However, one can clearly identify that the scaling
regime breaks down for large voting fractions and a faster decay takes
place. This breakdown of scaling invariance, not seen in data from
city-councillors elections, is similar to the one reported to be
present in linguistics \cite{linguistics} and re-association of folded
proteins \cite{proteins}. The crossover to a faster (exponential)
decay for large voting fractions can reflect a lack of correlation
among groups of voters that chose to vote for the same candidate but
are located in distinct cities. Within the reasoning that leads to
Eq.~(\ref{dif}), this crossover can be reproduced by considering that
$N(v)$ follows a more general equation \cite{proteins}
\begin{equation}\label{difext}
\frac{dN}{dv} = -\lambda N -(\lambda_q-\lambda)N^q~,
\end{equation}
whose general solution reads
\begin{equation}\label{Zipfext}
N(v)=\frac{N(0)}{[1-\lambda_q/\lambda +(\lambda_q/\lambda)
e^{{(q-1)\lambda v}}]^{1/(q-1)}}~.
\end{equation}
In Fig.~2 the solid line corresponds to the best fit of $N(v)$ to
Eq.~(\ref{Zipfext}). From it we estimate $\lambda_q=12227.3$, $\lambda
= 135.1$ and $\alpha=1/(q-1)=1.03$ ($q=1.97$). The exponent $\alpha$
is in perfect agreement with the previously estimated value.

The difference in scaling exponents appearing in the Zipf's law for
city and state representatives leads to the conjecture that local and
non-local voting processes have distinct underlying dynamics governing
the voting decision. For the city representative election, a model
based in the Sznajd model of sociophysics \cite{sznajd} has been
proposed to reproduce the exponent $\alpha\simeq 1.0$ \cite{bernardes}.
However, a model that captures some characteristics of non-local
voting processes would possible require the inclusion of complex
voting decision rules that take into account the role of parties,
economic and social factors as well as long-range interacting
individuals, among others. The natural question that arises is
therefore related to the possibility of a universal behavior of
non-local voting processes and what factors are relevant in defining
the scaling exponent.

In order to test for a possible universal behavior of non-local voting
processes, we collected data from the National Congress (federal
deputies) election results of 1998. Again, we just took the data
corresponding to the same 15 states previously considered. The
resulting distribution is reported in Fig.~3. Notice that a breakdown
of scaling invariance is also observed in this case for large voting
fractions. The best fit to the generalized Zipf's law
Eq.~(\ref{Zipfext}) is shown as a solid line and the fitting
parameters are $\lambda_q=6774.4$, $\lambda=33.1$ and $\alpha
=1/(q-1)=1.07$ ($q=1.93$). The similarity between the exponents
characterizing the intermediate power-law regime of both state and
federal deputies comes in favour of the conjecture that non-local
voting processes depict a universal scaling behavior. However, the
crossover to the exponential behavior takes place at distinct
(non-universal) voting fractions.

\section{Conclusion}

In summary, here we provide a statistical analysis for the results of
the Brazilian elections for state and federal deputies held in 1998,
and for city-councillors held in 2000. We show that all regimes of the
measured distributions $N(v)$ are well reproduced by a generalized
form of Zipf's law which has been originally derived within a
nonextensive thermostatistical approach. More precisely, we find that
the voting distribution for city representatives follows Zipf's law
with a characteristic exponent $\alpha\simeq 2.6$. Furthermore, the
Zipf's law representing the data for state and federal deputies
exhibits a slower decay with $\alpha\simeq 1.0$, followed by a
non-universal crossover to an exponential decay which reflects the
lack of persistent correlations at large voting fractions. We
therefore conclude that the distributions for local and non-local
voting processes are characterized by distinct scaling exponents. At
this point, it is important to emphasize that these exponents are
practically the same among all states analyzed. This suggests that
voting motivations are indeed similar for different states regardless
of economic, social and political factors. On the other hand, voting
motivations for local (city-councillors) and non-local (state and
federal deputies) voting processes may be quite distinct leading to the
observed qualitatively different results. It would be interesting to
have models of social behavior able to distinguish between the
presently reported scaling behavior of local and non-local voting
processes.

\acknowledgements{This work was supported by the Brazilian
funding agencies CNPq, FUNCAP, and CAPES}

\section{Figure Captions}

\noindent
{\bf Figure 1} - Double logarithmic plot of the voting distribution
for city-councillors. Data concern to the results of the Brazilian
proportional elections held in 2000 considering the 15 largest state
capitals. The solid line is the best fit to the Zipf's law,
Eq.~(\ref{Zipf}). The power-law scaling exponent is $\alpha=2.63$.

\noindent
{\bf Figure 2} - Double logarithmic plot of voting distribution for
state deputies. Data concern to the results of the Brazilian
proportional elections held in 1998 considering the states with the 15
largest capitals. The solid line is the best fit to the generalized
Zipf's law, Eq.~(\ref{Zipfext}). The power-law scaling exponent is
$\alpha=1.03$. The crossover to an exponential decay reflects the lack
of persistent correlations at large voting fractions. The scaling
exponent is much smaller than the one characterizing the
city-councillors election indicating that these two processes have
distinct voting decision rules.

\noindent
{\bf Figure 3} - Double logarithmic plot of the voting distribution
for federal deputies. Data concern to the results of the Brazilian
proportional elections held in 1998 considering the states with the 15
largest capitals. The solid line is the best fit to the generalized
Zipf's law, Eq.~(\ref{Zipfext}). The power-law scaling exponent is
$\alpha=1.07$. The similarity with the scaling exponent obtained for
state deputies indicates that these two process may be dominated by
the same voting decision rules.

\begin{figure}
\narrowtext
\epsfxsize=9.0cm\epsfysize=8.0cm
\centerline{ \epsffile{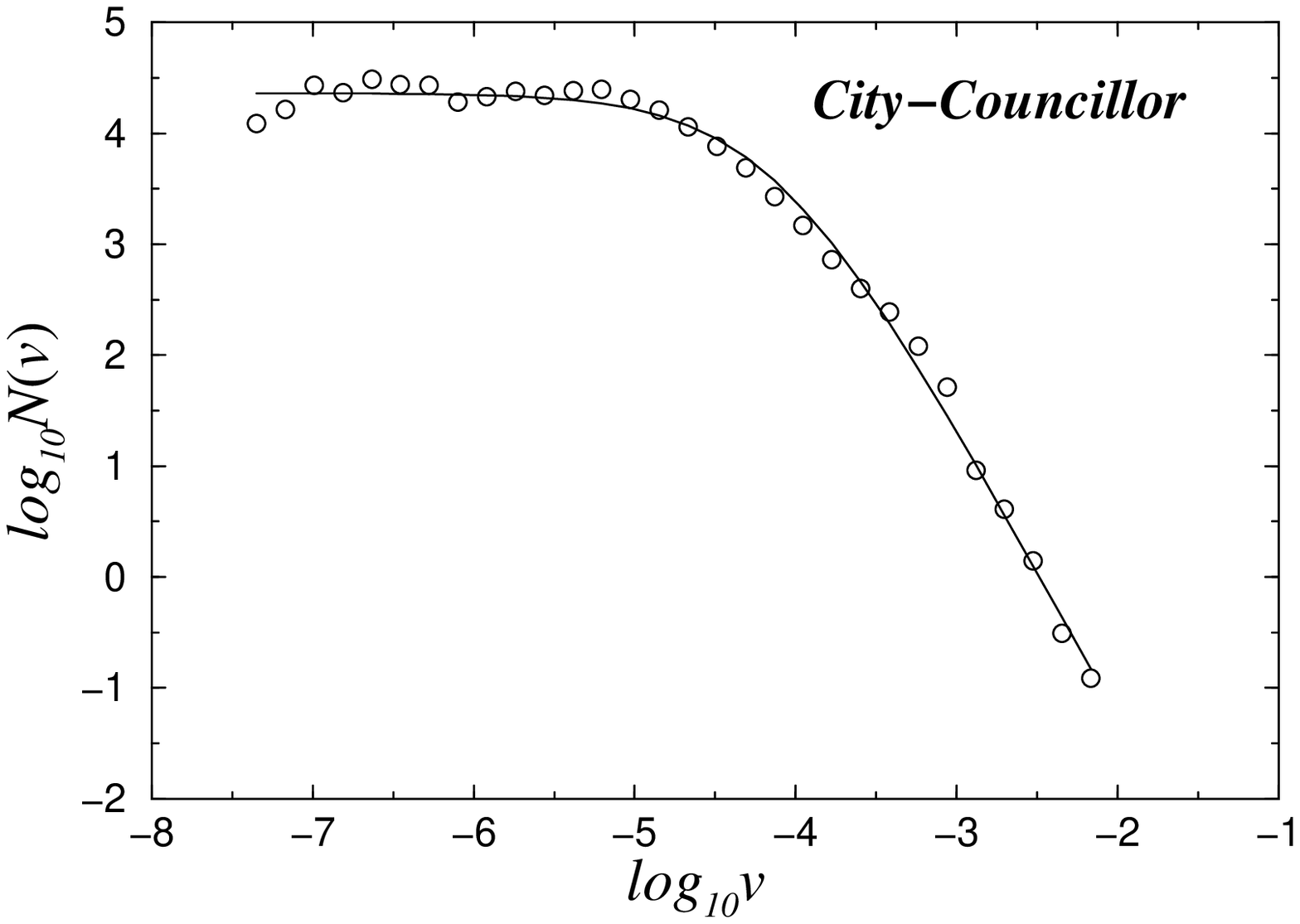} }
 
\caption{}
\end{figure}

\begin{figure}
\narrowtext
\epsfxsize=9.0cm\epsfysize=8.0cm
\centerline{ \epsffile{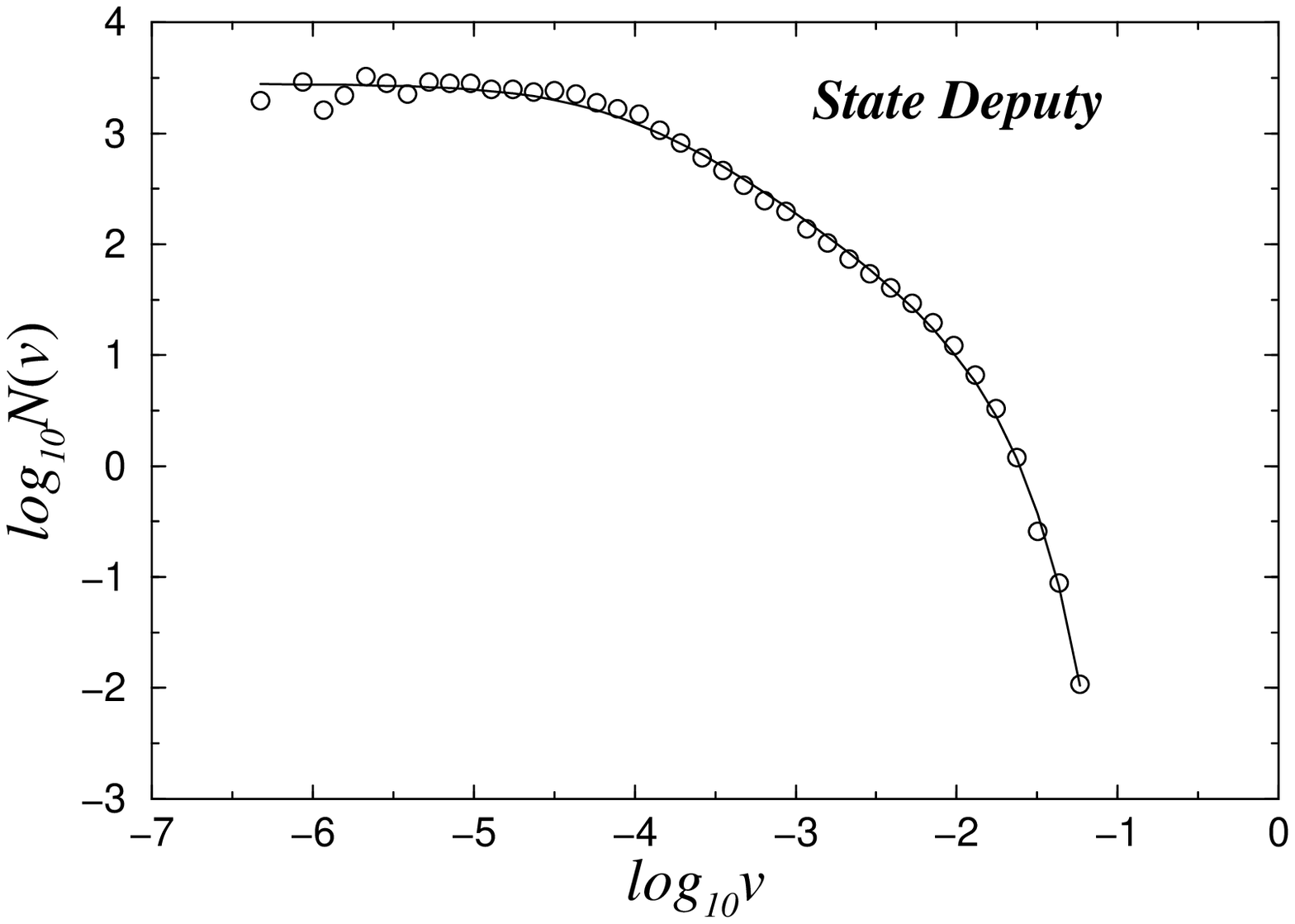} }

\caption{}
\end{figure}

\begin{figure}
\narrowtext
\epsfxsize=9.0cm\epsfysize=8.0cm
\centerline{ \epsffile{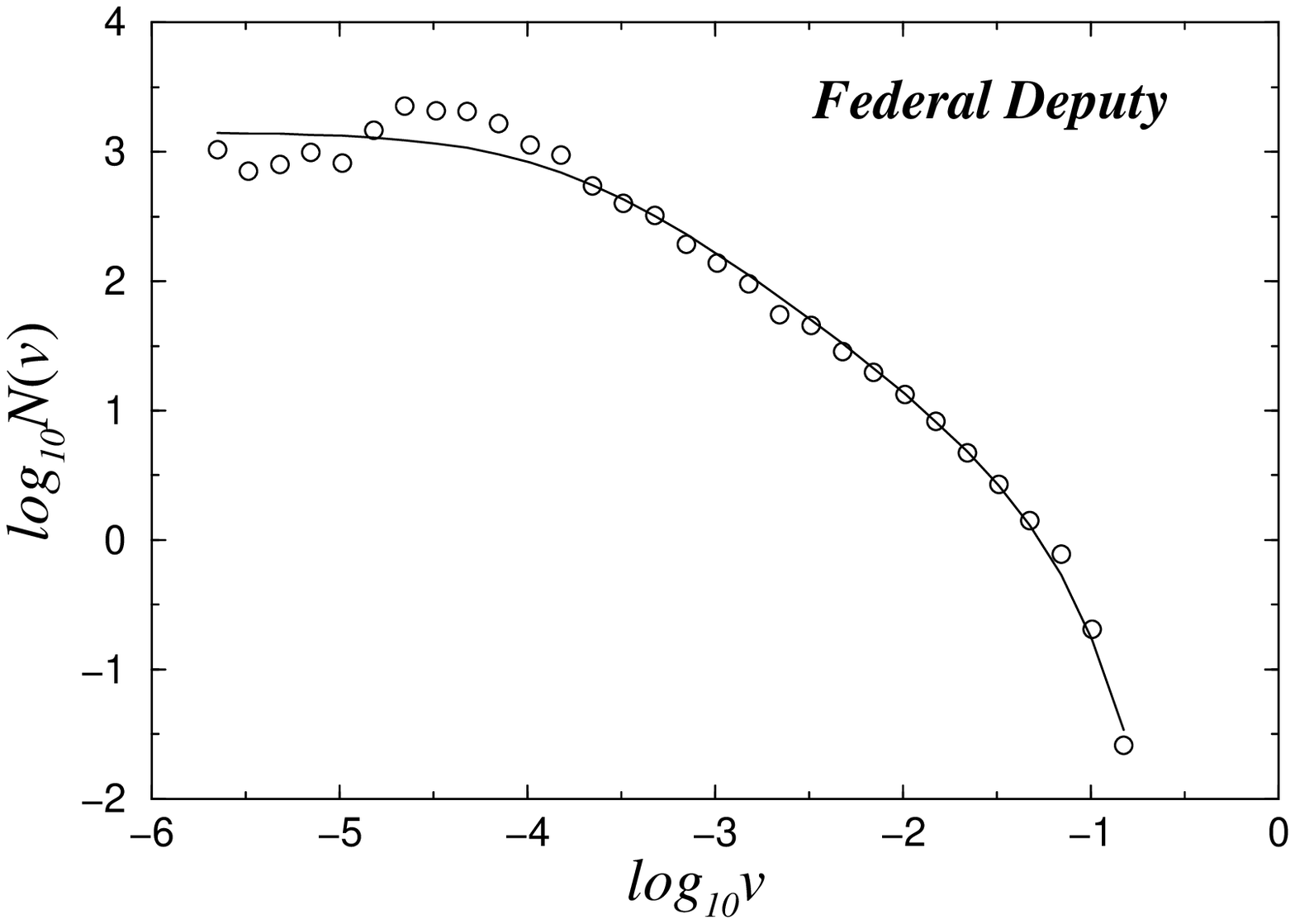} }

\caption{}
\end{figure}
\end{document}